# On the mechanism of generation of the Kelvin-Helmholtz instability

Alexander Ershkovich

*Department of Geophysics and Planetary Sciences, Tel Aviv University*

**Abstract.**

Mechanism of the Kelvin-Helmholtz instability based on the Kutta-Zhoukovsky theorem is considered. The mechanism itself generates the velocity shear, thereby redistributing the flow energy which drives the instability.

Kelvin-Helmholtz instability is of great interest in physics, astrophysics, geophysics, plasma physics and engineering [1-4]. The energy source of the instability is the kinetic energy of the flow redistributed by means of the velocity shear [5]. There are a number of mechanisms of shear instability generation, e.g. wave-interaction mechanism [6]. The most simple mechanism is considered below, with the scenario mentioned in [1], namely: a perturbation, say an inflation appears sporadically on the interface (a tangential discontinuity). The fluid velocity flowing around the inflation increases. According to Bernoulli law, the pressure decreases there, and the inflation will grow further (due to the pressure gradient ) being convected downstream. The same effect may arise in a uniform flow because the flow around the obstacle (a perturbation) itself generates the velocity shear. It is worth mentioning that the dispersion equation for a sheared flow with the transition layer of thickness $d$ reduces to that for a tangential discontinuity interface (called the vortex sheet approximation) for long wavelength perturbations with $kd \ll 1$ ([1, 5]).

Let us consider two-dimensional uniform potential flow of ideal incompressible fluid, and a small perturbation in the flow arising sporadically. According to Kutta-Zhoukovsky theorem, the force exerting on the planar rigid closed contour of any form in such a flow is (e.g. [7])

$$F = -i\rho u \Gamma \qquad (1)$$

where $u = v_\infty$ is the flow velocity at infinity, $\rho$ is the density, and $\Gamma = \oint \mathbf{v} d\mathbf{r}$ is the velocity circulation around the contour. The real part of the complex force $F = F_x + iF_y$ is zero, so that in ideal (inviscid) fluid the drag vanishes. This is so called d'Alembert paradox. There is a simple physical explanation of the formula (1) by means of the interference of two independent motions, translation and circulation. As a result of this interference the velocity shear arises. According to Bernoulli law, the pressure gradient appears, thereby generating the 'lifting' force (1). Of course, the rigid contour is a streamline, and in the steady state ($\partial/\partial t = 0$) flow streamlines coincide with particle trajectories.

Potential steady-state uniform flow (of ideal incompressible fluid) around a circular cylinder (a circle) is a useful and instructive example because any closed planar contour may be reduced to a circle by means of conformal mapping. The complex potential $W$ of the flow around a circle of radius $R$ is [8]

$$W(z) = uz + \frac{uR^2}{z} + \frac{\Gamma}{2\pi i}\ln z \equiv \Phi + i\Psi \qquad (2)$$

where $z \equiv x + iy = r\exp(i\theta)$. Two stagnation points $z_{1,2}$ are found with $dW/dz = 0$ and happen to occur at $r = R$ under the condition

$\Gamma^2 < (4\pi u R)^2$. In this case the velocity of the fluid particle, $\mathbf{v} = \nabla\Phi$ is

$$v = v_\theta = \frac{\Gamma}{2\pi R} - 2u\sin\theta \equiv b - c\sin\theta \qquad (3)$$

According to the hypothesis of the fluid continuum, two fluid particles, located near the stagnation point $z_1$, on the opposite sides of the stagnation streamline have to meet again near the stagnation point $z_2$ after the same time $\tau$. Let us check this prediction. According to (3), one obtains

$$\tau = \int_{\theta_1}^{\theta_2} \frac{R d\theta}{v(\theta)} = \frac{R}{\sqrt{c^2 - b^2}} \ln\left|\frac{b\tan\dfrac{\theta}{2} - c - \sqrt{c^2 - b^2}}{b\tan\dfrac{\theta}{2} - c + \sqrt{c^2 - b^2}}\right|_{\theta_1}^{\theta_2} \qquad (4)$$

with $\Gamma^2 < (4\pi u R)^2$, i.e. $b^2 < c^2$.

Indeed, the time $\tau$ depends only on locations $z_1$ and $z_2$ on the circle $r = R$, and does not depend on the direction of the particle rotation. Thus, particles will meet again. However the distances $z_1 z_2$ (along the circle $r = R$) are different for these two particles, and if time of their trip between the points $z_1$ and $z_2$ is the same, their average velocities are also different. This example proves the fact that the force proportional to the pressure gradient should arise. It is easy to find the force on the unit length of a cylinder (circle) by using equation (3). One obtains that $F_x = 0, F_y = -\rho u\Gamma$ and, hence, the complex force $F = -i\rho u\Gamma$, in accordance with equation (1).

It is worthy of mentioning that the hypothesis of the fluid continuum does not always hold, and particles moving together may not meet again. This rather often occurs, for instance, under fluid sprinkling, under cavitation, etc.

An example above is indicative of violation of Thomson (Lord Kelvin) theorem on conservation of the velocity circulation. Indeed, when rigid contour appears in a potential flow of ideal fluid the flow (at least at the contour vicinity) becomes rotational. This fact is described by a curl-term proportional to circulation Γ (which is a given parameter of the problem). How it may happen? Well, rigid contour is a streamline. According to continuum hypothesis, a streamline in a flow cannot remain empty. Thus, the fluid circulation should arise together with obstacle. The case Γ = 0, of course, cannot be eliminated but looks less probable. It is quite natural to assume tentatively that a sporadic perturbation in the flow, in this respect, is similar to a rigid contour. Then, if Γ ≠ 0 a small perturbation will grow under the action of 'lifting' Kutta-Zhoukovsky force proportional to Γ, resulting in the Kelvin-Helmholtz instability.

**Conclusion**

The most simple mechanism of the Kelvin-Helmholtz instability generation is considered. It is based on Kutta-Zhoukovsky law using a similarity between flow around both a rigid contour and a perturbation (which may sporadically arise at the vortex sheet discontinuity or in the homogeneous potential flow of ideal incompressible fluid). We arrive at the conclusion that non-zero velocity circulation (required for the instability to arise) should be generated together with a perturbation. The velocity shear which supplies the instability with the flow energy is generated by the mechanism under consideration.

**References**

1. Ershkovich, A., 1980, Space Sci. Rev., **25**, 3.

2. Ray, T. and Ershkovich, A., 1983, MNRAS, **204**, 821.

3. Ershkovich, A., and Israelevich, P., 2015, J. Plasma Phys., 81, 3, doi: 10.1017/S0022377814001275.

4. Ershkovich, A., and Israelevich, P., 2015, On the stability of cylindrical tangential discontinuity, generation and damping of helical waves, Cornell Univ. Library, http://arXiv.org/abs/1502.00989.

5. Chandrasekhar, S., 1961, Hydrodynamic and Hydromagnetic stability, Clarendon Press, Oxford, Chapter 11.

6. Baines, P., and Mitsudera, H., 1994, J. Fluid Mechanics, **276**, 327 (see also references therein).

7. Lamb, H., 1932, Hydrodynamics, Cambridge Univ. Press, ch. 11.

8. Batchelor, G., 1970, An Introduction to Fluid Dynamics, Cambridge Unversity Press, ch. 6.

4